\documentclass[conference]{IEEEtran}
\IEEEoverridecommandlockouts
\usepackage[sort,compress]{cite}
\usepackage{amsmath,amssymb,amsfonts}
\usepackage{algorithmic}
\usepackage{graphicx}
\usepackage{textcomp}
\usepackage{subfigure}
\usepackage{xcolor}
\usepackage{geometry}
\def\BibTeX{{\rm B\kern-.05em{\sc i\kern-.025em b}\kern-.08em
    T\kern-.1667em\lower.7ex\hbox{E}\kern-.125emX}}
\begin{document}
\title{A Privacy-Preserving Trajectory Synthesis Method Based on Vector Translation Invariance Supporting Traffic Constraints\\

\thanks{This work was supported by the National Natural Science Foundation of China (No. 62372340 and No. 62072349), and Major Technical Research Project of Hubei Province (No. 2023BAA018).}
}

\author{\IEEEauthorblockN{Zechen Liu\textsuperscript{1}, Wei Song\textsuperscript{1,2*}, Yuhan Wang\textsuperscript{1}}
\IEEEauthorblockA{
\textit{\textsuperscript{1}School of Computer Science, Wuhan University, Wuhan, China}\\
\textit{\textsuperscript{2}College of Information Science and Technology, Shihezi University, Shihezi, China}\\
\textit{\textsuperscript{*}Corresponding Author: songwei@whu.edu.cn}\\
}
}

\maketitle

\begin{abstract}

With the popularization of different kinds of  smart terminals and the development of autonomous driving technology, more and more services based on spatio-temporal data have emerged in our lives, such as online taxi services, traffic flow prediction, and tracking virus propagation. However, the privacy concerns of spatio-temporal data greatly limit the use of them. To address this issue, differential privacy method based on spatio-temporal data has been proposed. In differential privacy, a good aggregation query can highly improve the data utility. But the mainstream aggregation query methods are based on area partitioning, which is difficult to generate trajectory with high utility for they are hard to take time and constraints into account. Motivated by this, we propose an aggregation query based on the relationships between trajectories, so it can greatly improve the data utility as compared to the existing methods. The trajectory synthesis task can be regarded as an optimization problem of finding trajectories that match the relationships between trajectories. We adopt gradient descent to find new trajectories that meet the conditions, and during the gradient descent, we can easily take the constraints into account by adding penalty terms which area partitioning based query is hard to achieve. We carry out extensive experiments to validate that the trajectories generated by our method have higher utility and the theoretic analysis shows that our method is safe and reliable.
\end{abstract}


\section{Introduction}
Spatial-temporal data based services have long been widely used with the popularization of GPS technology and various location-aware devices. With the advent of the big data era, the issue of users' data privacy has become more and more serious. The traditional $k$-anonymization and $l$-diversity algorithms can no longer guarantee the privacy of users\cite{DBLP:journals/tkde/JinHFCOZ23,klt,attackKL,attackKL2}. Researchers have adopted Differential privacy\cite{Dwork06}, which has better privacy protection effect, to protect users' privacy during data services. However, the privacy protection of user data using differential privacy requires the construction of aggregation queries. A good aggregation query can help the algorithm to reduce the global sensitivity and thus improve the data utility. For the spatio-temporal data, mainstream aggregation queries are based on area partitioning \cite{DPT,ngram,areaLimitGridBase,PrivTrace,TMC,clusterPub}. This kind of methods always scatter the sample points of a trajectory into areas and then perturb the sample points number within an area. In synthesis part, these methods adopt random walk methods based on perturbed possibility they captured in original dataset to reconstruct the trajectories.
Such methods have a number of disadvantages. For instance, they are difficult to perturb the temporal nature of trajectories thus they can not thoughtfully protect privacy. For examples,\cite{AdaTrace,DPT,ngram,PrivTrace,TMC} under the area partitioning based method, the time corresponding to the starting point and the ending point of a certain user is difficult to perturb with differential privacy\cite{TMC}, and many papers even do not perturb the time directly\cite{PrivTrace,ngram,DPT}. What's more, the effect is strongly influenced by the partition granularity. Fine-grained region partitioning and coarse-grained partitioning will not only directly affect the amount of noise, but also the selection of trajectory points during trajectory synthesis. Fine-grained partitioning will make the perturbed trajectory look more realistic, but more noise. And coarse-grained partitioning can reduce the noise, but the perturbed trajectory will be very unrealistic, with only a few simple folds\cite{TMC,PrivTrace}. These can all greatly affect the usability of the published data. Moreover, different maps (dataset) have different optimal granularity \cite{ngram,DPT,DPstar}. However, there are no papers presenting demonstrable methods that can find the optimal granularity.
Area partitioning based methods are difficult to meet the constraints. In real-world, many areas and roads are closed at certain times of the day. If we directly generate trajectories without take those constraints into account. It will generate unrealistic trajectories which greatly reduce the usability of the synthetic dataset. Some methods have been proposed, such as \cite{areaLimitGridBase,beihang}, to eliminate these unrealistic perturbations under the area partitioning based method, but the efficiency is low. Unrealistic trajectories or points will be discarded when generating the trajectories and then regenerate trajectories or points. As described in \cite{areaLimitGridBase}, the generation may involve $50,000$ iterations to ensure that all the trajectories have met the realistic constraints.
From this, it can be seen that area partitioning based methods are hard to efficiently generate trajectory dataset with high utility. To solve the problems listed above, \cite{PrivTrace} and \cite{TMC} try to optimize both the granularity of the area partition and the random walk algorithm as much as possible, so that the synthetic trajectory dataset is more similar to the original trajectory dataset in terms of the distribution of the trajectory points. \cite{beihang} and \cite{areaLimitGridBase} try to eliminate the unrealistic perturbation by divide reachable and unreachable area. But there is still no exploration for the problem of temporal, impractical perturbations.

To solve those problems and inspired by another Word2Vec algorithm (Random Walk algorithm is also used to achieve Word2Vec\cite{rdwalkW2Vec}), we propose a novel trajectory synthesis method named \textbf{DPE} (\textbf{D}ifferential \textbf{P}rivacy \textbf{E}mbedding). We regard trajectory as a high-dimensional sequence, not only latitude and longitude but also \textbf{TIMESTAMP}. And we construct a metric space over the set of trajectories and an aggregation query by the measure between the trajectories. We protect trajectory data by perturbing the metrics between trajectories. There's no need to find the optimal granularity. The trajectory synthesis problem is transformed into finding a new set of trajectories which can fit the perturbed metrics.

In contrast to the existing random walk based algorithms \cite{AdaTrace,TMC,PrivTrace,LDPTrace,prefixtree} which utilize Markov train, etc., this problem can be regarded as an optimization problem of minimizing $d(T')-d(T)$, where $d$ represents the value of the metric and $T,T'$ represent the original trajectory dataset and the perturbed trajectory dataset respectively.
The optimization target at synthesis step are more explicit, and various constraints can be added as penalty terms $P_i$ after the optimization function $d(T')-d(T)+\sum \mu_i P_i$, where $\mu_i$ denotes the coefficient of penalty terms. Search-based methods, iterative random walk based methods can be avoided to generate realistic trajectories. See section.\ref{method} for the specific method.

Of course, the method of considering the whole trajectory as a whole to be perturbed has been used by \cite{shipDP}, but the it did not utilize the inter-trajectory relationship of the whole dataset to construct the aggregation query which causes a huge amount of noise and reduces the data utility.

Our main contributions can be summarized bellow:
\begin{itemize}
    \item We propose a new type of aggregation query which can take temporal nature into account. 
    \item The proposed approach \textbf{DPE} is able to add various penalty terms so that it can express various realistic limitations. Unlike other methods, our method does not need to be repeated many times to avoid the limitations. The gradient of the penalty term can be utilized to avoid the restricted region during the gradient descent method. 
    \item Our method can generate more realistic trajectories efficiently, the experimental results show that our method is much more faster and the trajectories generated by our method have better utilization.
\end{itemize}

\section{Related works}

\subsection{Differential Privacy Trajectory Synthesis}
Publishing sensitive trajectory data with differential privacy is becoming the mainstream method for its superb privacy-preserving capability. At the time when differential privacy is just being used to trajectory publishing, researchers are coming up with a myriad of frameworks. For examples, \cite{shipDP,shipdpwithDTW} directly perturb the coordinates of the trajectory, but these methods directly use the diameter of a map as the coordinate's or trajectory's global sensitivity. This will cause synthetic dataset contains very large noise and has very low usability. Other researchers turn the trajectory into a sequence of place names, replacing the coordinates with labels such as "XX store", "XX school", etc. Then the Laplace mechanism or exponential mechanism is utilized to protect privacy \cite{safepath}. Some other methods divide the map into several grids or regions, then count the number of sampling points in each region to generate a heatmap. They extract the probabilistic features of dataset from this heatmap, and then use the perturbed heatmap and perturbed probabilistic features to generate the synthetic trajectories. Because the global sensitivity of each grid/region is $1$. The global sensitivity of this kind of methods is smaller than that in the first method. After \cite{DPT,ngram}, most of the area partitioning based methods are tuned for sampling and generating trajectories according to probability features. However, these methods erase the continuity and temporal features of the trajectory by dividing them into grids, which inevitably reduces the usability of data in another way. So far, the usability of spatio-temporal data after privacy-preserving processing still can not satisfy the needs in the real-world applications. Moreover, the existing privacy-preserving synthesis methods are difficult to generate trajectories that meet the constraints. \cite{areaLimitGridBase,beihang} tried to filter out the trajectories which do not meet the constraints during the generation process, but their methods have to spend many iterations to filter out the trajectories which do not meet the constraints.

In order to address the common drawbacks of these methods and to improve the usability of the synthetic dataset, we propose an aggregation query based on the relationships between trajectories in this paper, which utilizes \emph{Vector Translation Invariance} to generate trajectories.
\subsection{Trajectory Synthesis}
Currently, most of the mainstream area partitioning methods adopt $n$-gram-like algorithms \cite{ngram,DPT,PrivTrace,TMC,LDPTrace} to synthesize trajectory. In $n$-gram-like algorithms, each location is regarded as a "word", and then find out the most likely "word" after this "word". The disadvantages of these methods mentioned above are difficult to be solved. So we made a new attempt to apply another mainstream method of Word2Vec, distance-based embedding method \cite{TransE,TransH,TransG}, to trajectory generation. The first distance-based embedding method, TransE\cite{TransE} model in 2013, found that there is a translation invariant phenomenon in the word-vector relationship. And this is fully compatible for the trajectory space modeled with Hilbert space. But in a given trajectory dataset, the dimension of trajectories is constant. So, we construct a vector space with specific metric to embed the trajectories into the new space using the translation invariance phenomenon among trajectories. Then, we use the relationships between the trajectories to construct the aggregation query. Although there are many improved variants of TransE model later such as TransH\cite{TransH}, TransG\cite{TransG}, etc., their motivation is mainly the translation invariant phenomenon of word-vector relations does not conform to the strict definition of distance. But the metric of trajectories in vector space is consistent, so we only draw on the original TransE model.
\section{Preliminary}
\subsection{Differential Privacy}
Differential Privacy has a strong privacy guarantee for it ensures that the output of a private algorithm is not strongly dependent on any one trajectory in the input dataset.\\
\textbf{Definition 1:}(Neighbouring Dataset) \emph{We call $\mathcal{D}$ and $\mathcal{D}'$ neighbouring dataset when $\mathcal{D}=\mathcal{D}'\cup\left\{T\right\}$ or $\mathcal{D}'=\mathcal{D}\cup\left\{T\right\}$ which means $\mathcal{D}$ and $\mathcal{D}'$ only have one different record.}\\
\textbf{Definition 2:}($\varepsilon$-Differential Privacy) \emph{An algorithm $\mathcal{M}$ satisfies $\epsilon$-differential privacy, when $\mathcal{M}$ satisfies constraints below:}
\begin{equation}
    \forall \mathcal{O} \subset \mathcal{R}ange\left(\mathcal{M}\right):Pr\lbrack\mathcal{M}\left(\mathcal{D}\right)\in\mathcal{O}\rbrack\leq e^\varepsilon Pr\lbrack\mathcal{M}\left(\mathcal{D}'\right)\in\mathcal{O}\rbrack
    \label{def of dp}
\end{equation}
\emph{where $\mathcal{R}ange\left(\mathcal{M}\right)$ denotes the output domain of randomize algorithm $\mathcal{M}$}.\\
\textbf{Laplace Mechanism}: Laplace Mechanism is a randomize algorithm which satisfies $\varepsilon$-differential privacy by adding random noise to the output of query function $f$ on dataset $\mathcal{D}$. The final answer $\mathcal{A}$ to the query $f$ can be expressed by the equation below:
\begin{equation}
    \mathcal{A}_f\left(\mathcal{D}\right) = f\left(\mathcal D\right) + Lap\left(\frac{\lambda}{\varepsilon}\right) 
\end{equation}
where $\lambda$ denotes the global sensitivity of dataset $\mathcal{D}$ on query $f$. \emph{Global Sensitivity can be defined as the equation below:}
\begin{equation}
    \lambda = \max \left\| f\left(\mathcal{D}\right)-f\left(\mathcal{D}'\right)\right\|
\end{equation}
$Lap\left(\frac{\lambda}{\varepsilon}\right)$ denotes the random noise sampled from the Laplace distribution. When $f$ outputs a vector, the mechanism adds noise to each element of the vector.\\
\textbf{Post-Processing:} Given an $\varepsilon$-differential privacy mechanism $\mathcal{M}$, any post process does not reduce the privacy guarantee. This has been proved by Dwork in \cite{Dwork06}.
\subsection{Problem Statement}
We separate the publishing task into two parts: aggregation query design and optimal synthesis dataset finding.\\
\textbf{Aggregation Query Design:} Given a trajectory dataset $\mathcal{D}$ with $m$ trajectories, the aggregation query $q$ of the $i$-th trajectory $T_i$ which we designed for $\mathcal{D}$ is defined as the equation below:
\begin{equation}
    q\left(\mathcal D,T_i\right) = \sum\limits_{j=1,i\neq j}^m d\left(T_i,T_j\right)
    \label{query definition}
\end{equation}
where $d$ denotes the metric of trajectories $T_i$ and $T_j$. We assume that every trajectory dataset has its own scope $\mathcal{S}$. For example, the scope of Geolife\footnote{www.microsoft.com/en-us/research/publication/geolife} dataset is Beijing because the research focuses on Beijing and its trajectories are all sampled in Beijing.\\
\textbf{Synthesis Dataset:} We model the synthesis procedure as an optimal problem. Given the perturbed query result $\tilde q$ for each trajectory in dataset $\mathcal{D}$. Our task is to find the trajectories $T'$ in scope of  $\mathcal{D}$ which meet the perturb query result $\tilde q$.
\begin{equation}
    \arg\min\limits_{T \subset \mathcal{S}} q\left(\mathcal{D}',T'_i\right)-\tilde q\left(\mathcal{D},T_i\right)
    \label{task}
\end{equation}
\textbf{Constraint:} A constraint is an expressible area in trajectory space $U_T$ in which the trajectories do not meet the real-world conditions. Let $D_\alpha$ denotes the real-world dataset samples from a specific region at a time. We call an area $\mathcal{S}_c$ constraint when it satisfies the conditions below:
\begin{equation}
    \mathcal{S}_c \in U_T, D_\alpha \in U_T;
\end{equation}
\begin{equation}
    \forall T_i \in \mathcal{S}_c, T_i \notin \bigcup\limits_{\alpha}^\infty D_\alpha
\end{equation}
\section{Method}
\label{method}
As shown in Fig.1, our method DPE is divided into four steps: trajectory embedding, embedding space transformation, trajectory perturbation, and trajectory generation. We draw on the graph embedding method TransE\cite{TransE}. \textbf{DPE} embeds the trajectory into a metric space and calculates the metric sum between the trajectory and other trajectories. Then DPE perturbs the metric sum of the trajectory using the Laplace mechanism. Finally, DPE adds penalty terms to Eq.\eqref{task} according to the constraints. We then obtain the synthetic trajectory dataset by re-embedding according to the perturbed metric sum.
\begin{figure*}[htbp]
\label{fig:structure}
    \centering
    \includegraphics[width=0.8\linewidth]{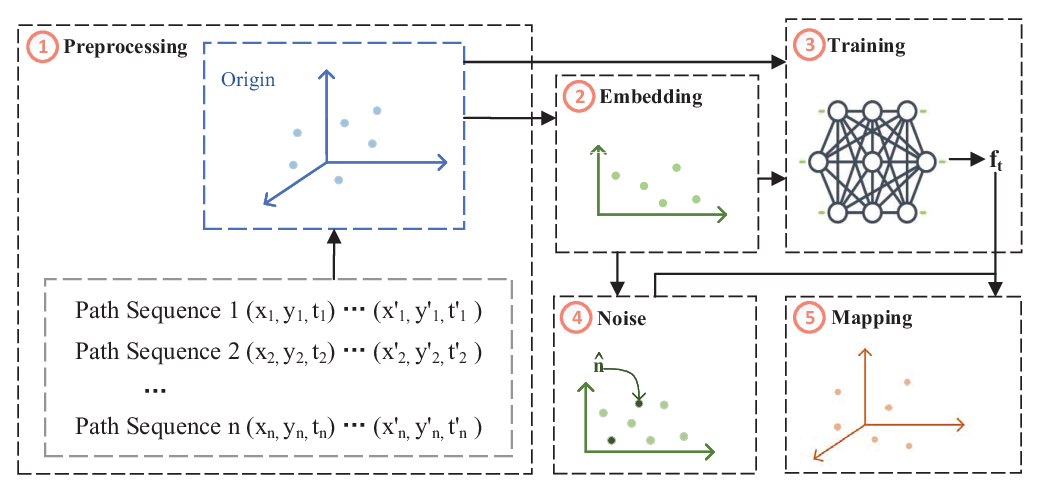}
    \caption{Overview of DPE. We first represent the trajectories as a high-dimensional vector and embed them into another metric space with a particular metric. Secondly, we train a function $f_t$ which can map the embedded trajectory to original trajectory. Thirdly, we re-embed the trajectory with perturbation and obtain the perturbed embedded trajectories. At last, we map the perturbed trajectories to original trajectory space with $f_t$ trained in second step.}
\end{figure*}

\subsection{Trajectory Embedding}
A spatio-temporal trajectory is a sequence in three-dimensional space, which can be represented as a series of ${x,y,t}$ where $x$ denotes latitude, $y$ denotes longitude, and $t$ denotes timestamp. We find that $x,y,t$ are linear in a given region (e.g., a city) and the value domains are $\mathbb{R}$. By the definition of linear space, we know that a trajectory can be viewed as a point in a $3n$-dimensional linear space, where $n$ denotes the number of points contained in a trajectory. We then construct the metric in this space to make it a metric space.

There are numerous methods for computing the metric between trajectories, such as Fréchet Distance, Euclidean Distance, and so on. Due to the requirement of gradient descent method, as long as the metric is first order derivable, it can be used as the metric of our method. In this paper, we choose Euclidean distance as an example.

	Since the Euclidean distance requires that the lengths of the trajectories are all the same, we use the first and last extension method to align the trajectories. That is, the trajectory is considered to have been stopped in the place before the beginning and after the end. We choose the longest trajectory in the dataset as the benchmark trajectory, then calculate the length difference $l$ between the trajectory and the benchmark trajectory. We copy $l/2$ times of the start and end points of the trajectory respectively, and splice them at the beginning and the end of the trajectory respectively to form the aligned trajectory.

Different metrics require different pre-processing. Since this is not the focus of this paper, we skip it in this article. The presentation continues using the Euclidean distance as an example.

Once the pre-processing operations required by the metric are completed, we select the dimension $D'$ of the space to be embedded, and then randomly generate an initial set of post-embedding trajectories $\tilde T'$. We regard embedding task as an optimization problem. That is to find a set of $T'$ which can minimize the Eq.\eqref{target function}. We adopt gradient descent to select the optimal $T'$ to complete the embedding.
\begin{equation}
    \mathcal{L}\left(T'\right) = \sum_i^n\left(\sum_j^n d_e\left(T_i,T_j\right)-d_e\left(T'_i,T'_j\right)\right)
    \label{target function}
\end{equation}
where $d_e$ denotes the Euclidean distance. 
If $D'$ is equal to the dimension of the original space $D$, then the embedding process here is just doing a spatial transformation, and the relationship information among the trajectories is completely preserved. If $D' < D$, then the embedding process can downsize the trajectories, which reduces the amount of computation but will result in a loss of relationship information. $D'$ can be larger than $D$, but there is no need to be larger than that. Of course, when $D'=D$, it is also possible to just use the original trajectories as the initial post-embedding trajectories $\tilde T'$ which saves the next step.

Unlike the traditional partition granularity, $D'$ here has a well-defined optimal value in terms of accuracy ($D'\geq D$), which is not affected by the map, but only affected by the user's computational resource. Thus we can better guarantee the usability of the final synthetic trajectory dataset.
\subsection{Embedding Space Transformation}
After embedding trajectories into a embedding space, we need to establish a mapping between the embedding space and the original space. We consider the mapping task as a fitting problem. If embedding dimension $D'$ is equal to original dimension $D$, we can directly use the original trajectories as the initial trajectories $\tilde T'$ in the previous step, then this step is not needed. Otherwise, if $D' \neq D$, then we need to fit a function $f_t$ which can map the trajectory in embedding space to the corresponding trajectory in original space. We call the $f_t$ transformation function. For this function, we adopt the linear function for fitting. $f_t$ can be expressed as follows:
\begin{equation}
    f_t\left( T'_i\right) = \omega T'_i+b
    \label{transformation function}
\end{equation}
where $T_i$ and $T'_i$ means the $i$-th trajectory in original trajectory space and embedding space respectively, $\omega$ denotes the weights, $b$ denotes the bias.

\subsection{Trajectory Perturbation}
In this paper, we use the Laplace mechanism for perturbation. As shown by Eq.\eqref{query definition}, we take the metric sum of each trajectory with other trajectories $d_{s}$ as an aggregation query, and add noise to the metric sum of each trajectory with other trajectories. The formula is as follows:
\begin{equation}
    \tilde{d_s} = d_s + Lap \left(\frac{\lambda}{\varepsilon}\right) 
\end{equation}
where $\varepsilon$ is the privacy budget and $\lambda$ is the global sensitivity. In a trajectory dataset, the global sensitivity is determined by the time span as well as the boundary of the map in which the trajectory dataset is located. We define that the diameter of a map is the distance between the two farthest points from each other on the boundary of that map. For a trajectory dataset, we assume that its corresponding map has a diameter of $2r$ and a time span of $\tau$. Then $\lambda$ for this trajectory dataset can be calculated by equation bellow as.
\begin{equation}
    \lambda = \sqrt{n\cdot\left(4\cdot r^2+\tau^2\right)}
\end{equation}
where $n$ denotes the number of points contained in the longest trajectory.
So the overall perturbation formula is:
\begin{equation}
    \tilde{d_s} = d_s + Lap \left(\frac{\sqrt{n\cdot\left(4\cdot r^2+\tau^2\right)}}{\varepsilon}\right)
    \label{perturbation equation}
\end{equation}

\subsection{Trajectory Synthesis}
After we get the metric and $\tilde{d_s}$ of each trajectory after perturbation. We generate the perturbed trajectories based on these perturbed metrics and the embedded trajectory dataset $T'$ obtained by the embedding process. We take the  $\tilde{d_s}$ of each trajectory and the embedded trajectory dataset $T'$ into Eq.\eqref{task} to re-do the embedding. After the perturbed embedded trajectory dataset $T'_p$ is obtained. We take $T'_p$ as input into the function $f_t$ to obtain the perturbed trajectory dataset $T_p$ in the original space.\\

\textbf{With Constraints:} For example, a certain area with boundary $\mathcal{S}_c$ is off-limits during the time period $t_1$ and $t_2$. We brought it into the formula $f_t$ to obtain the boundary $\mathcal{S}'_c$ in embedding space. Then we can express the penalty term of this limitation by the formula $\mathcal{R}eLU\left(d_e\left(point,\mathcal{S}'_c\right)\right)$. The output of this formula will greater than $0$ if the point in $\mathcal{S}_c$. We denote it as $\mathcal{P}$ and add it to Eq.\eqref{target function}:
\begin{equation}
    \mathcal{L}'\left(T'\right) = \sum_i^n\left(\sum_j^n d_e\left(T_i,T_j\right)-d_e\left(T'_{pi},T'_{pj}\right)\right) + \sum_k^m \mathcal{P}_k
    \label{target function with penalty}
\end{equation}

According to Lagrange theorem, as long as the penalty factors are properly chosen, the constraints can be avoided during gradient descent \cite{nipsPenalty} without the need for repeating the generation over and over again until a trajectory is generated like grid partitioning based and random walk based methods such as \cite{areaLimitGridBase,beihang}.

Moreover, the grid partitioning method can't represent the constraints such as speed limit, one-way line, etc., but our method can represent them by penalty terms. It makes our generated trajectory more close to the actual trajectory and has better utility. It can also satisfy the $\varepsilon$-differential privacy requirements.

\section{Privacy Analysis}
Our proposal satisfies $\varepsilon$-differential privacy. According to the definition of differential privacy. Let $\mathcal{S}$ denotes the scope of dataset $\mathcal{D}$ and its neighbouring dataset $\mathcal{D}'$. The output $\mathcal{O}$ of query function $q$ on dataset $\mathcal{D}$ with the given input $T_i$ can be expressed as the equation below:
\begin{equation}
    \mathcal{O} = \sum\limits_{T_j\in\mathcal{D},i\neq j}^m d_e\left(T_i,T_j\right)
    \label{origin q on D}
\end{equation}
The output $\mathcal{O}'$ of query function $q$ on dataset $\mathcal{D}'$ with the same given input $T_i$ can be expressed as the equation below:
\begin{equation}
    \label{origin q on adjacent D}
    \mathcal{O}' = \sum\limits_{T_j\in\mathcal{D}',i\neq j}^m d_e\left(T_i,T_j\right)
\end{equation}
Because $\mathcal{D}$ and $\mathcal{D}'$ have only one different trajectory. We denote the different trajectory as $T^*$. The maximum $\left \|\mathcal{O}-\mathcal{O}'\right \|$ is the max mod of $T^*$. We denote the max mod of $T^*$ as $d^*$. According to the definition of $\varepsilon$-differential privacy and the global sensitivity, we have $\lambda = d^*$. In order to prove that our algorithm satisfies $\varepsilon$-differential privacy, we have to prove the following equation:
\begin{equation}
    Pr\lbrack q\left(T_i,\mathcal{D}\right) +
    Lap\left(\frac{\lambda}{\varepsilon}\right)\in\mathcal{O}\rbrack \leq e^\varepsilon Pr\lbrack q\left(T_i, \mathcal{D}'\right)+Lap\left(\frac{\lambda}{\epsilon}\right)\rbrack
    \label{proof target}
\end{equation}
We assume that the output $\mathcal{O}$ ranges from $s_1+q\left(T_i,\mathcal{D}\right)$ to $s_2+q\left(T_i,\mathcal{D}\right)$, we have:
\begin{align}
\label{Pro of noised q1}
    Pr\lbrack q\left(T_i,\mathcal{D}\right)+
    Lap\left(\frac{\lambda}{\varepsilon}\right) \in\mathcal{O}\rbrack &= \int_{s_1}^{s_2} Lap\left(\frac{\lambda}{\varepsilon}\right)\\
    Pr\lbrack q\left(T_i, \mathcal{D}'\right)+Lap\left(\frac{\lambda}{\varepsilon}\right)\in\mathcal{O}\rbrack &= \int_{s_1\pm\lambda}^{s_2\pm\lambda} Lap\left(\frac{\lambda}{\varepsilon}\right)
    \label{Pro of noised q2}
\end{align}
Because the formula of Laplace distribution is:
\begin{equation}
    Lap\left(\frac{\lambda}{\varepsilon}\right) = \frac{1}{2\lambda}e^{-\frac{\varepsilon|x|}{\lambda}}
    \label{Lap distribution}
\end{equation}
Equation.\eqref{Pro of noised q1} and Eq.\eqref{Pro of noised q2} can be rewritten into:
\begin{align}
\label{pro of q1}
    \int_{s_1}^{s_2} Lap\left(\frac{\lambda}{\varepsilon}\right) &=\int_{s_1}^{s_2} \frac{1}{2\lambda}\cdot e^{-\frac{\varepsilon |x|}{\lambda}}\\
    \int_{s_1\pm\lambda}^{s_2\pm\lambda} Lap\left(\frac{\lambda}{\varepsilon}\right) &= \int_{s_1\pm\lambda}^{s_2\pm\lambda} \frac{1}{2\lambda}\cdot e^{-\frac{\varepsilon |x|}{\lambda}} 
    \label{pro of q2}
\end{align}
Take Eq.\eqref{pro of q1} and Eq.\eqref{pro of q2} into Eq.\eqref{def of dp} we have:
\begin{equation}
    \int_{s_1}^{s_2} e^{-\frac{\varepsilon \lbrack x\rbrack}{\lambda}} \leq e^\varepsilon \cdot \int_{s_1\pm\lambda}^{s_2\pm\lambda}  e^{-\frac{\varepsilon \lbrack x\rbrack}{\lambda}} 
    \label{imeq}
\end{equation}

According to the properties of $0$-mean Laplace distribution, when output $\mathcal{O}$ has a fixed length, which means $|s_1-s_2|$ is stable, $\int_{s_1}^{s_2} e^{-\frac{\varepsilon |x|}{\lambda}}$ reaches the maximum value when $s_1 = -s_2$. Because $s_1\pm\lambda$ and $s_2 \pm\lambda$ can reach all $\mathcal{R}ange\left(\tilde q\right)$ as $\lambda$ can take any value. We assume that $s_2 \geq 0$ and $s_1 = -s_2$. The maximum value of $\int_{s_1}^{s_2} e^{-\frac{\varepsilon |x|}{\lambda}}$ equals $\frac{2\lambda}{\varepsilon}\left(1-e^{-\frac{\varepsilon\cdot s_2}{\lambda}}\right)$. As for $\int_{s_1\pm\lambda}^{s_2\pm\lambda}  e^{-\frac{\varepsilon |x|}{\lambda}}$, we discuss it in two circumstances. \\
\begin{figure*}[htbp]
\begin{center}
\footnotesize
\begin{tabular}{ccc}
\includegraphics[scale=0.38]{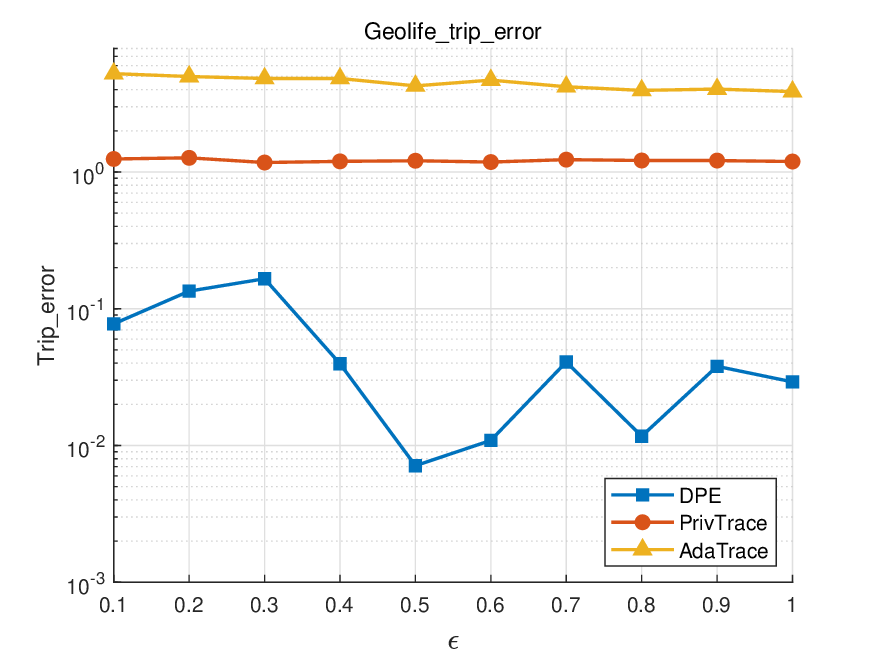}
\includegraphics[scale=0.38]{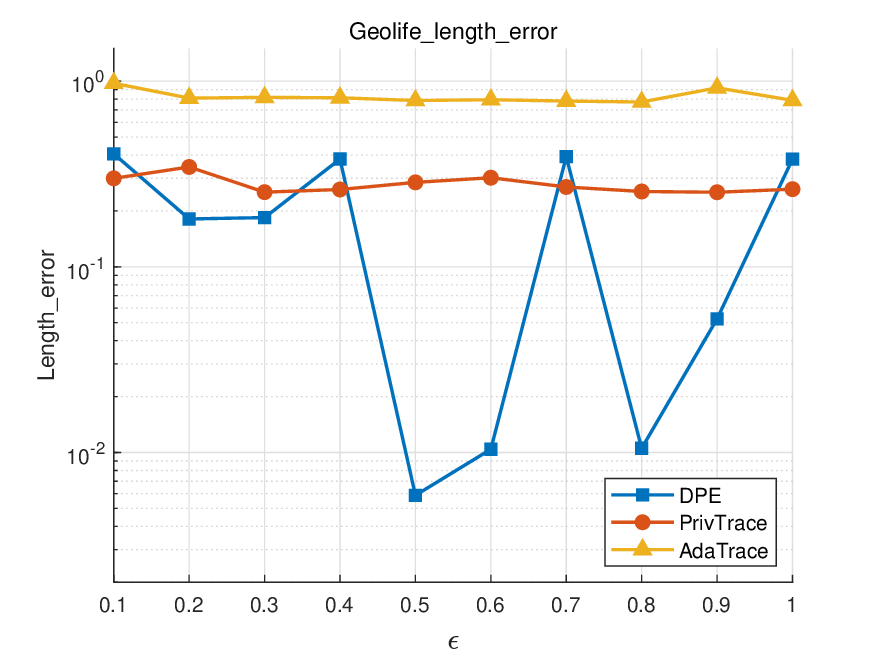}
\includegraphics[scale=0.38]{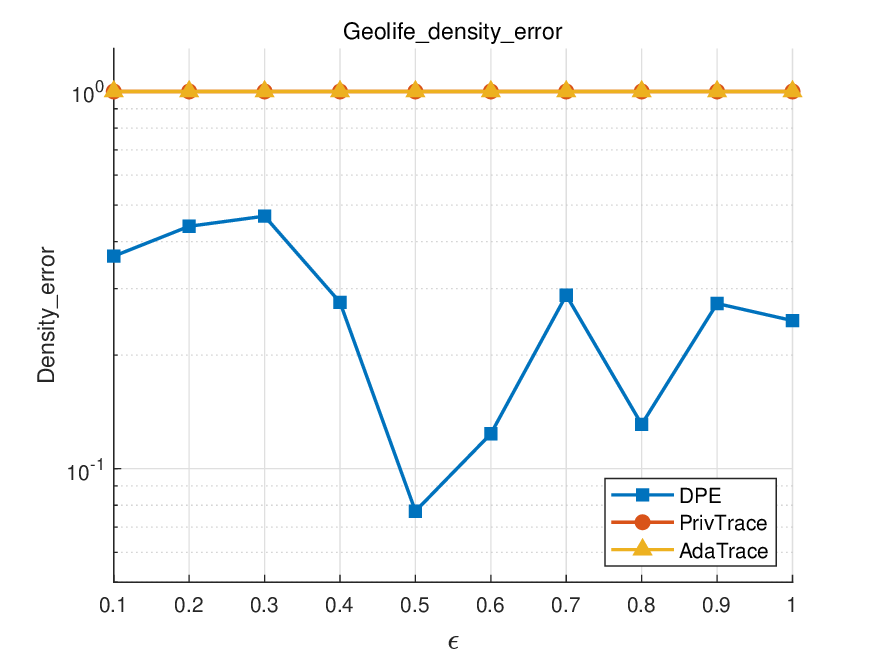}
\\
Geolife\\
\includegraphics[scale=0.38]{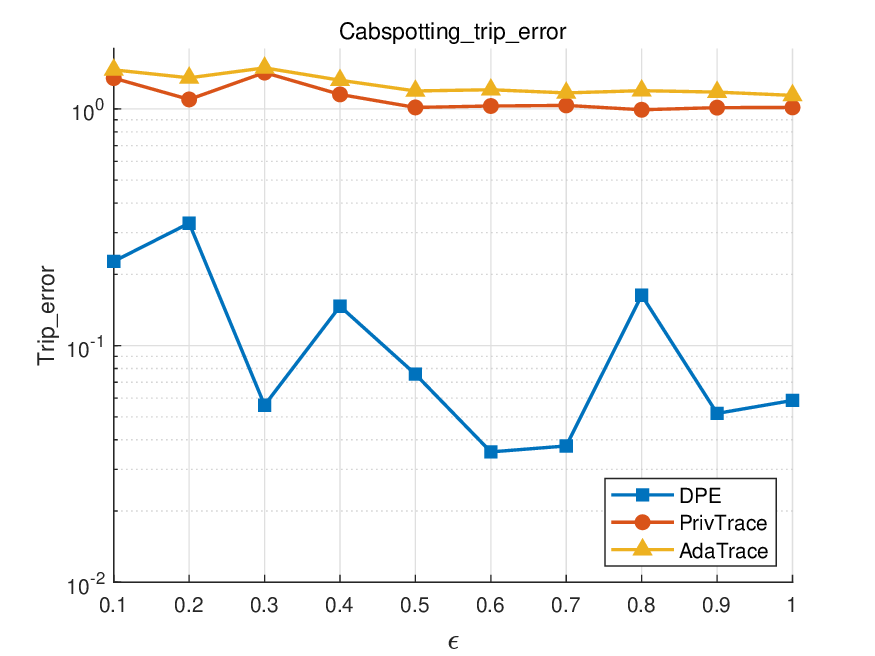}
\includegraphics[scale=0.38]{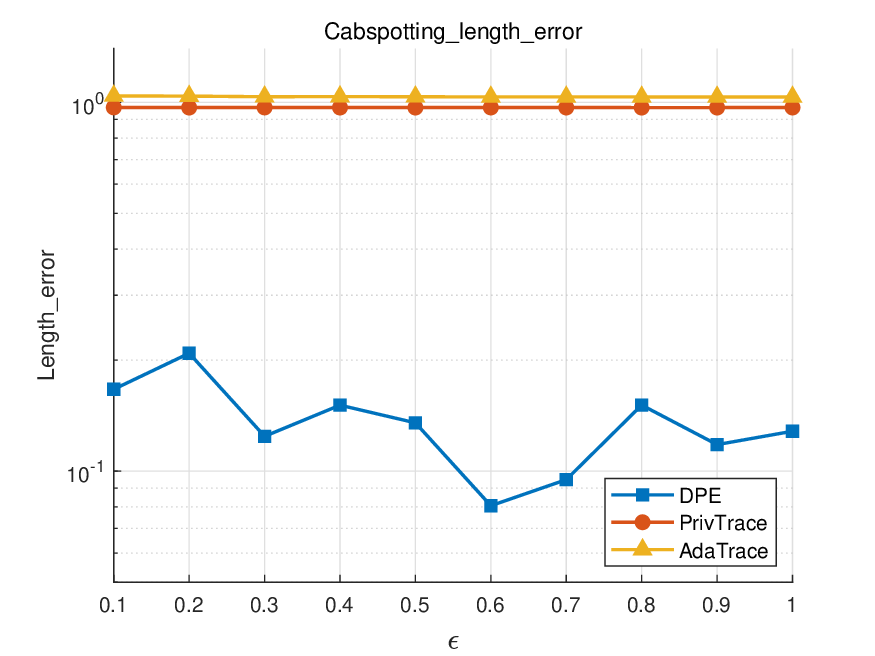}
\includegraphics[scale=0.38]{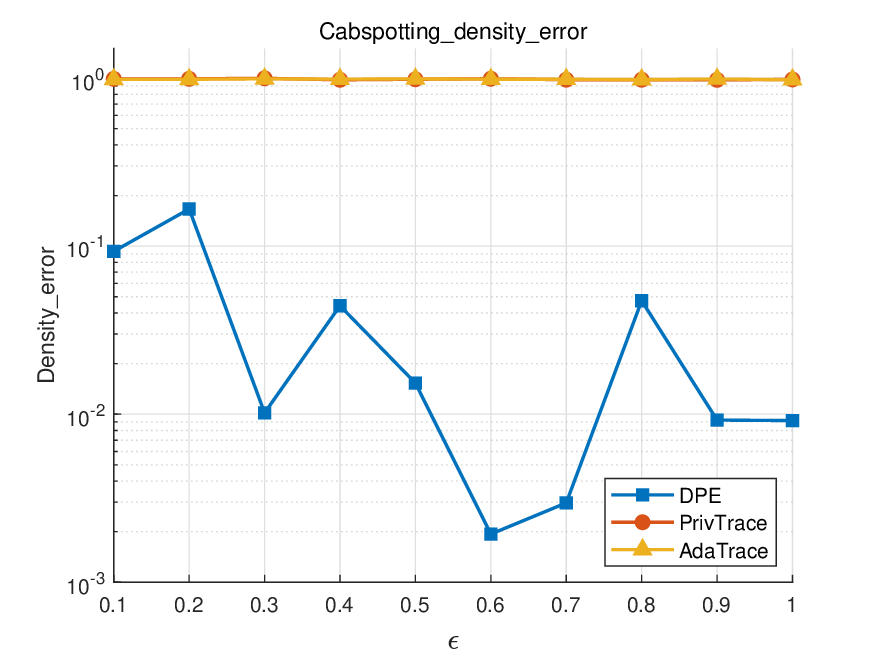}
\\
Cabspotting\\
\includegraphics[scale=0.38]{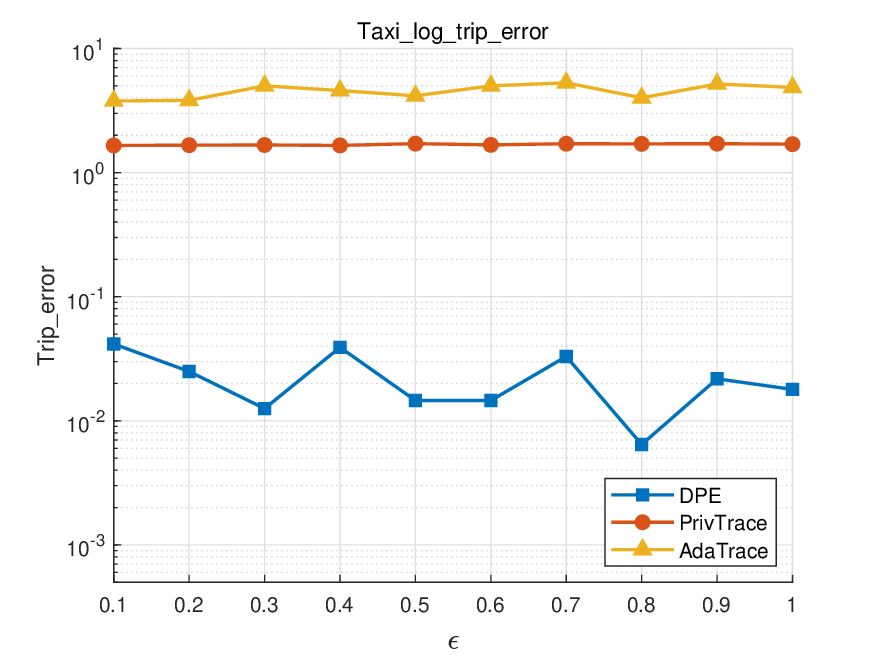}
\includegraphics[scale=0.38]{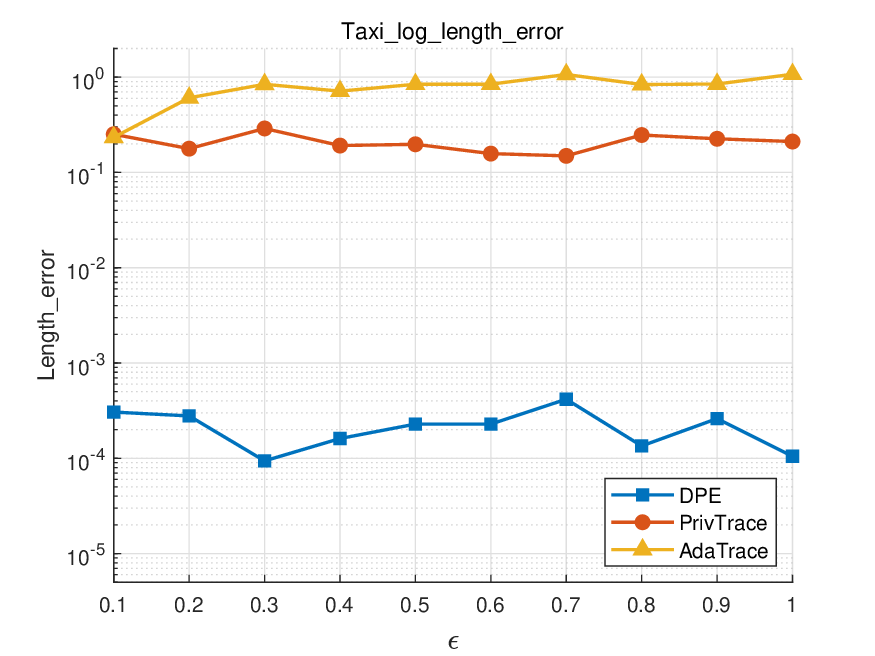}
\includegraphics[scale=0.38]{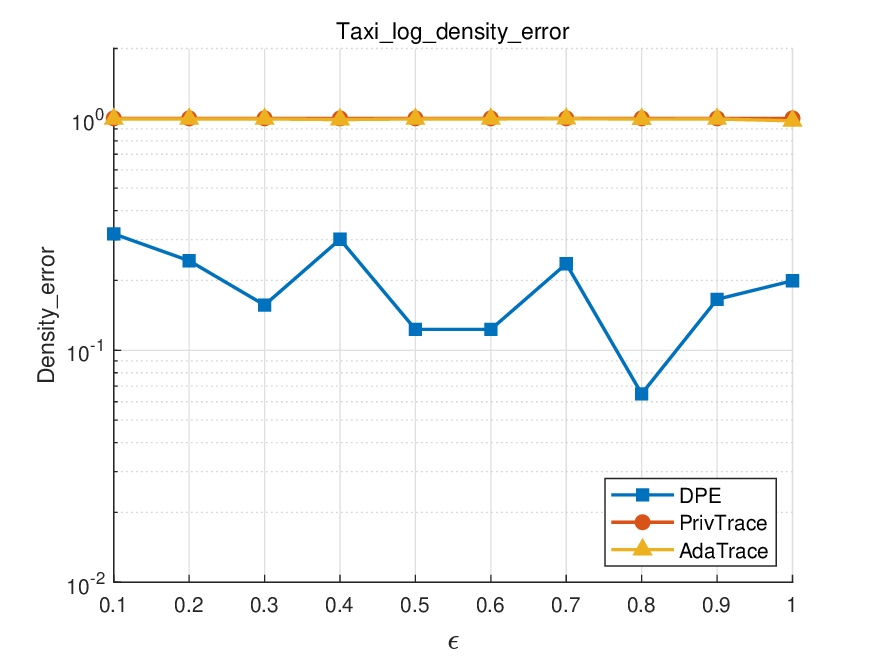}
\\
Taxi\\
\end{tabular}
\end{center}
\caption{Comparison Experiments}
\label{fig:ComparisonEx}
\end{figure*}
$\left(s_1\pm\lambda\right)\cdot\left(s_2\pm\lambda\right)\geq 0$: Because of the symmetry of $0$-mean Laplace distribution, we only consider $\left(s_1\pm\lambda\right)$ and $\left(s_2\pm\lambda\right)$ is beyond $0$. Take $\frac{2\lambda}{\varepsilon}\left(1-e^{\frac{\varepsilon\cdot s_2}{\lambda}}\right)$ and $\int_{s_1\pm\lambda}^{s_2\pm\lambda}  e^{-\frac{\varepsilon |x|}{\lambda}}$ into Eq.\eqref{imeq}, we have:
\begin{equation}
    \frac{2\left(1-e^{-\frac{\varepsilon\cdot s_2}{\lambda}}\right)}{\left(e^\varepsilon\right)^{\frac{s_2}{\lambda}\pm 1}-\left(e^\varepsilon\right)^{\frac{-s_2}{\lambda}\pm 1}}\leq e^\varepsilon
    \label{nearly imeq}
\end{equation}

When we take the value of $+\lambda$, and let $e^\epsilon = \beta$, Eq.\eqref{nearly imeq} becomes:
\begin{equation}
    \frac{2\left(1-\beta^{-\frac{s_2}{\lambda}}\right)}{\beta^2\cdot\left(\beta^{\frac{s_2}{\lambda}}- 1\right)+\beta^2\cdot\left(1-\beta^{\frac{-s_2}{\lambda}}\right)} \leq 1
\end{equation}
Because $\beta > 1$ and $s_2$ are greater than $0$, $\beta^{\frac{s_2}{\lambda}}- 1 \geq 1-\beta^{\frac{-s_2}{\lambda}}$. The inequality holds.
When we take the value of $-\lambda$, and let $e^\varepsilon = \beta$, Eq.\eqref{nearly imeq} becomes:
\begin{equation}
    \frac{2\left(1-\beta^{-\frac{s_2}{\lambda}}\right)}{\left(\beta^{\frac{s_2}{\lambda}}- 1\right)+\left(1-\beta^{\frac{-s_2}{\lambda}}\right)} \leq 1
\end{equation}
Because $\beta > 1$ and $s_2$ are greater than $0$, $\beta^{\frac{s_2}{\lambda}}- 1 \geq 1-\beta^{\frac{-s_2}{\lambda}}$. The inequality holds.\\

$\left(s_1\pm\lambda\right)\cdot\left(s_2\pm\lambda\right)\leq 0$: According to the monotonicity and symmetry of $0$-mean Laplace distribution. When the interval $|s_1-s_2|$ is fixed, the integral value reaches the minimum value when $s_1 \to 0$ or $s_2 \to 0$. Now we assume that $s_2 \pm \lambda = 2s_2$. Take it into Eq.\eqref{imeq}, we have:
\begin{align}
    \frac{2\left(1-e^{-\frac{\varepsilon\cdot s_2}{\lambda}}\right)}{1-e^{-\frac{\varepsilon}{\lambda}\left(2s_2\right)}} \leq e^\varepsilon
    \label{imeq of 2s}
    \\
    \frac{2\left(1-e^{-\frac{\varepsilon\cdot s_2}{\lambda}}\right)}{1-e^{-\frac{\varepsilon\cdot s_2}{\lambda}}+e^{-\frac{\varepsilon\cdot s_2}{\lambda}}-e^{-\frac{\varepsilon}{\lambda}\left(2s_2\right)}} \leq e^\varepsilon\\
    \frac{2}{1+e^{-\frac{\varepsilon\cdot s_2}{\lambda}}} \leq e^\varepsilon
\end{align}
Because $s_2\pm\lambda = 2s_2$, $e^{-\frac{\varepsilon\cdot s_2}{\lambda}} > 1$. The inequality holds.

We come to the conclusion that our proposal satisfies $\varepsilon$-differential privacy.

\section{Experiments}

In order to prove the effectiveness of our method, we design three experiments. (1) \textbf{Comparison experiment}, we choose state-of-the-art, PrivTrace\cite{PrivTrace} and AdaTrace\cite{AdaTrace} as our baseline, and compare our method with theirs in terms of length density error, trip error and density error. (2) \textbf{Parameter analysis experiment}, we analyze the performance of the model under different parameters as well as to prove the correctness of our method. (3) \textbf{Constraints experiment}, we simulate several restriction regions and design some penalty terms to prove that our method can indeed implement the controllable perturbation to meet constraints.

\subsection{Comparision Experiments}
\subsubsection{Experimental Setting}
We selected three dataset \textbf{Cabspotting}\footnote{crawdad.org/epfl/mobility/20090224},\textbf{T-drive}\footnote{www.microsoft.com/en-us/research/publication/t-drive-trajectory},\textbf{Geolife}\footnote{www.microsoft.com/en-us/research/publication/geolife} to test DPE, PrivTrace\cite{PrivTrace} and AdaTrace\cite{AdaTrace}. We implement our method with python3.9 on Windows11. We download the code of PrivTrace and AdaTrace from their public code shared in their papers. We run the experiment with their default settings. In this experiment, we set $D'= D$, and learning rate of each dataset is $0.1,0.05,0.8$ respectively. We adjust the privacy budget from $0.1$ to $1.0$ with step $0.1$. 
\subsubsection{Metrics}
We evaluate the utility of our method and PrivTrace,AdaTrace with length density error, trajectory density error and trip error. \\
\textbf{Length Density Error:}The length of a trajectory measures the summation of distances between all two adjacent points. We divide the length into $400$ bins from $0$ to the max length and count the number of trajectories falling into each bin to calculate the length distribution. We use the Jensen-Shannon divergence\cite{jsd} (JSD) to measure the error between original dataset and synthetic dataset.
\\
\textbf{Trajectory Density Error:} We divide the area of a dataset into cells and count the number of sample points within each cell. Every cell is $100m*100m$. We calculate the density of each cell and compute the density distribution. We use 
 Jensen-Shannon divergence \cite{jsd}(JSD) to measure the error between original dataset and synthetic dataset.\\
\textbf{Trip Error:} Trip error\cite{prefixtree} is to quantify the preservation of start/end cells for each trip, which is defined as the grid-based Jensen-Shannon divergence between trip distributions of original and synthetic dataset. Every cell is $100m*100m$ too.\\
\subsubsection{Result}
As shown in Fig.\ref{fig:ComparisonEx}. Our method has better performance in trip error, trajectory density error, length density error which demonstrate that our method has better utility. 

\subsection{Parameter Analysis}
\begin{figure}[htbp]
\centering
\subfigure[MSE Error of Heatmap]{
\begin{minipage}[t]{0.45\linewidth}
\centering
\includegraphics[width=1.8in]{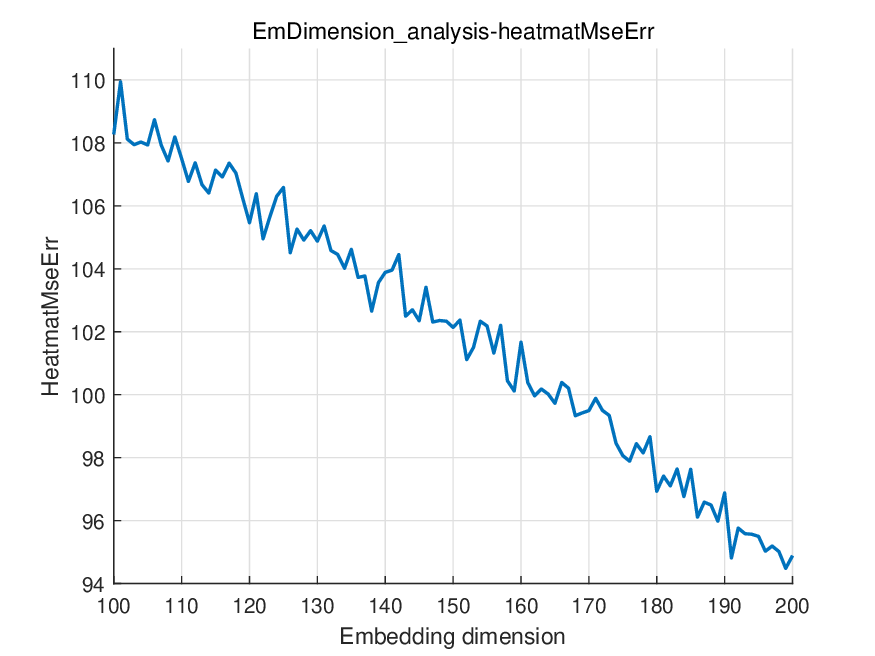}
\end{minipage}%
}%
\subfigure[MSE Error of Trajectories]{
\begin{minipage}[t]{0.45\linewidth}
\centering
\includegraphics[width=1.8in]{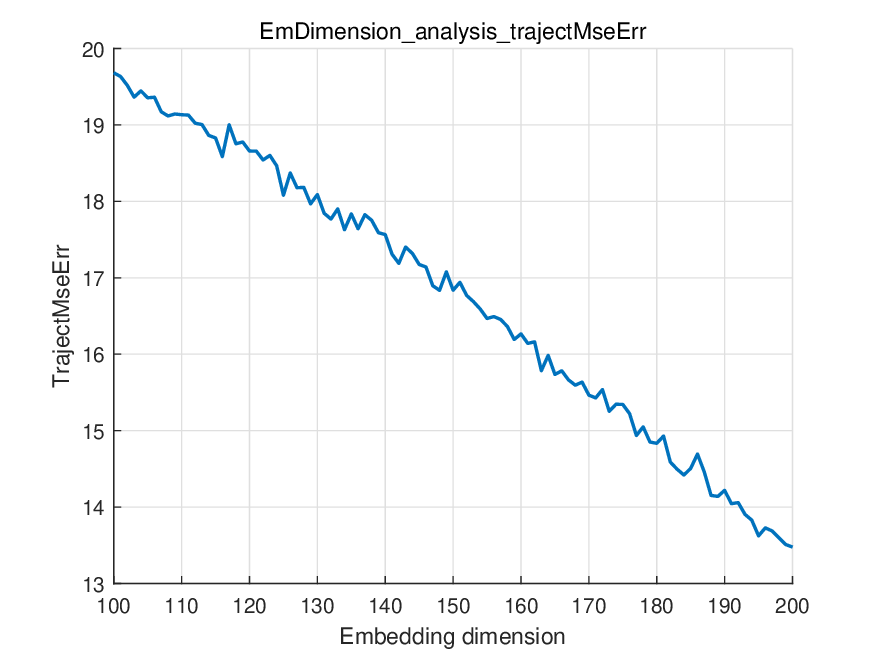}
\end{minipage}%
}%
\\
\centering
\subfigure[MSE Error of Heatmap]{
\begin{minipage}[t]{0.45\linewidth}
\centering
\includegraphics[width=1.8in]{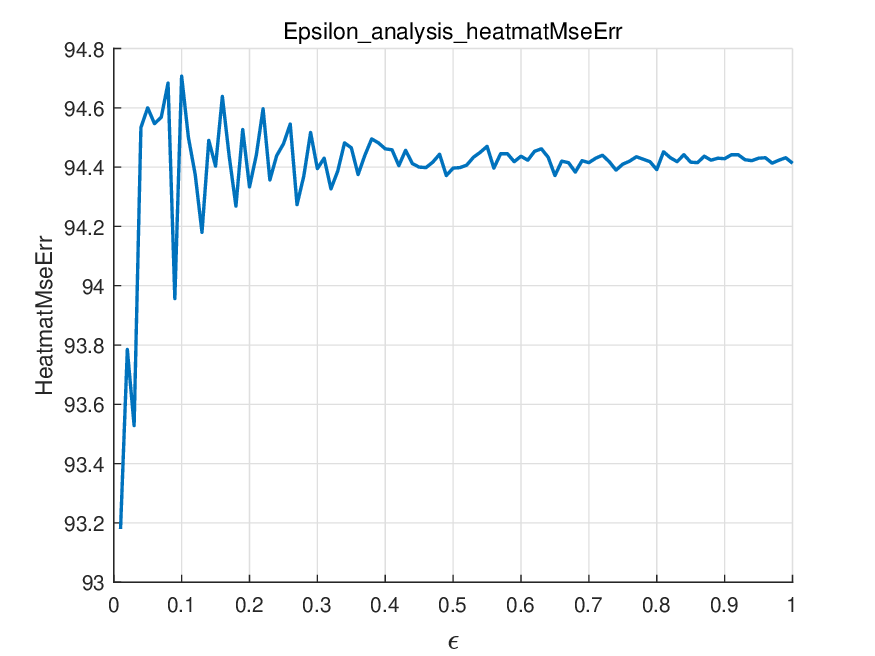}
\end{minipage}%
}%
\subfigure[MSE Error of Trajectories]{
\begin{minipage}[t]{0.45\linewidth}
\centering
\includegraphics[width=1.8in]{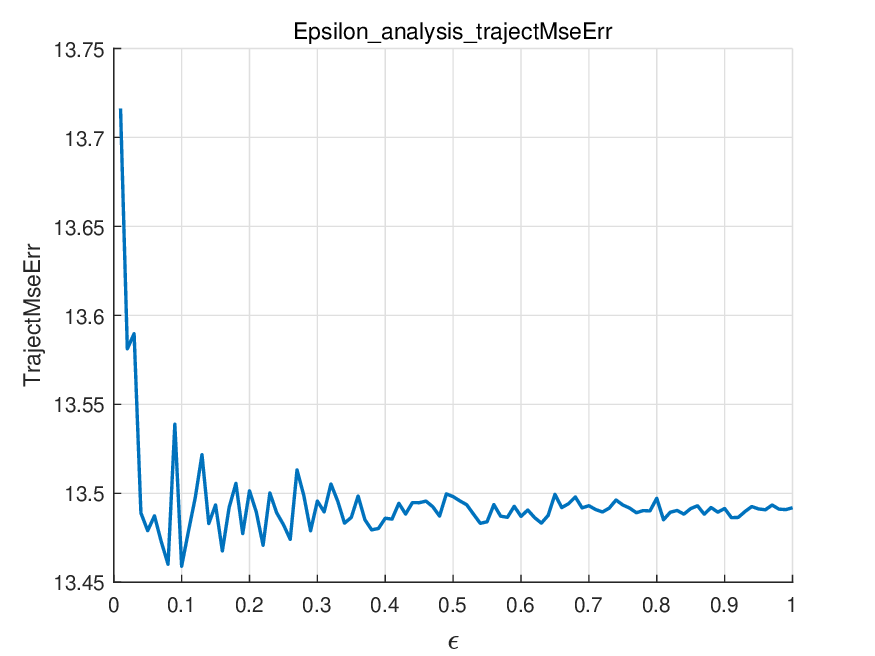}
\end{minipage}%
}%
\\
\centering
\caption{Parameter Analysis. We analyze the trend of model performance with changes in embedding dimension $D'$ and epsilon $\varepsilon$.}
\label{fig:ParamAnalysis}
\end{figure}

To analysis the impact of $D'$, we generate a simulated dataset with $500$ trajectories. The length of simulated trajectory is $100$. The scope of simulated dataset ranges from $\left(-10,-10\right), \left(-10,10\right)$ to $\left(10,-10\right), \left(10,10\right)$. We adjust $D'$ from $200$ to $100$ and $\varepsilon$ from $0.01$ to $1$ with step $0.01$. We use average MSE loss to quantify the error of synthetic trajectories. We set learning rate of embedding as $0.005$. The maximum number of iterations is $500$. As shown in Fig.\ref{fig:ParamAnalysis}, when $D'$ goes up, the error becomes smaller in proportional, which demonstrate that the theoretical analysis of our paper is correct and reveal that the optimal $D'$ is equal to the original $D$. 

\begin{figure}[htbp]
\centering
\subfigure[Without penalty terms]{
\begin{minipage}[t]{0.5\linewidth}
\centering
\includegraphics[width=1.5in]{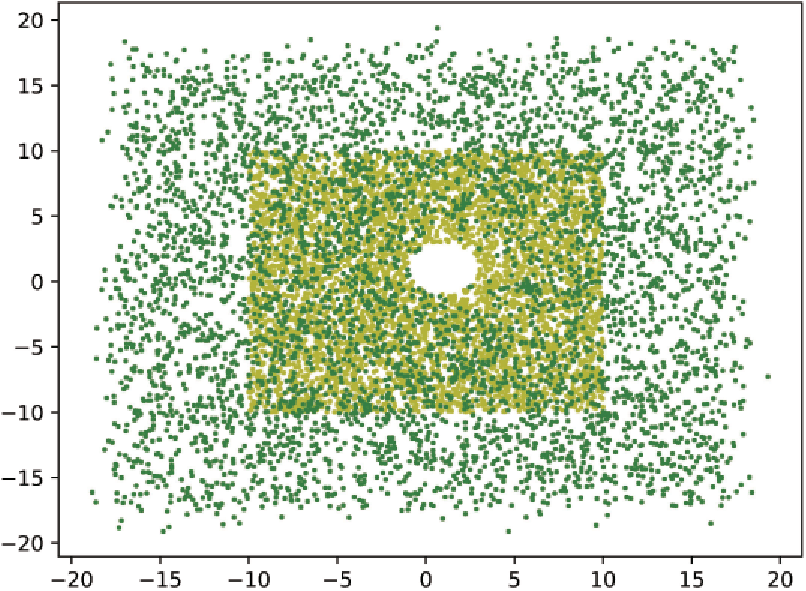}
\end{minipage}%
}%
\subfigure[With penalty terms]{
\begin{minipage}[t]{0.5\linewidth}
\centering
\includegraphics[width=1.5in]{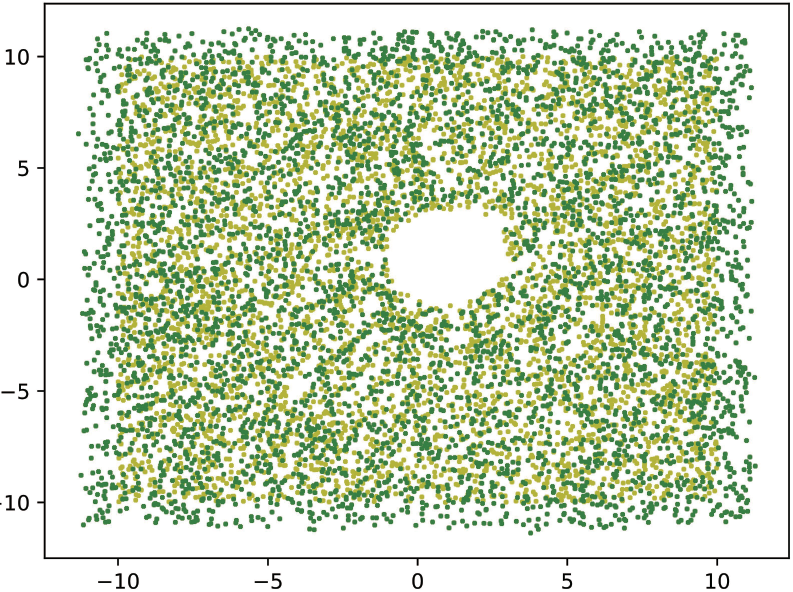}
\end{minipage}%
}%
\centering

\caption{Constraints Experiments:The yellow points are from original trajectories while green points are from perturbed trajectories. We can see that the one with penalty terms can avoid restricted area.}
\label{fig:RestrictionEx}
\end{figure}

\subsection{Constraints Experiment}
To demonstrate the ability of our method to adapt to constraints. We generate a simulated dataset using the same method as parameter analysis but we set $ x^2 + y^2 < 4 $ as restricted region. And we need to keep the perturbed trajectories lie in $\left(-10,-10\right), \left(-10,10\right)$ to $\left(10,-10\right), \left(10,10\right)$. We use two groups of synthesis dataset to evaluate the ability. One has add penalty terms while another is not. We adopt trajectory density to reveal the ability of adaptation to constraints. We set $\varepsilon=0.01, lr=0.1$. The penalty factors of circle and boundary are $0.3,5$ respectively. And we generate $50$ simulated trajectories. As shown in Fig.\ref{fig:RestrictionEx}, the group with penalty term avoid the restricted region while another group not which demonstrate that our proposal has ability to avoid restricted region.

\section{Conclusion}
In this paper we proposed a novel privacy-preserving trajectory publishing method with DP guarantees. By adopting TransE model to trajectory embedding and aggregation query construction, we achieve higher utility than state-of-the-art method PrivTrace. The algorithm introduction and privacy analysis is discussed in this paper and the experimental result shows that our method has the ability to adapt to the constraints while other methods do not.

\bibliographystyle{IEEEtran}
\bibliography{references}
\vspace{12pt}
\end{document}